\documentclass{article}

\usepackage{microtype}
\usepackage{graphicx}
\usepackage{subfigure}
\usepackage{booktabs} 
\usepackage{xcolor}

\usepackage{hyperref}

\hypersetup{
    colorlinks=true,
    linkcolor=promptblue,
    urlcolor=promptblue,
    citecolor=promptblue
}

\usepackage[accepted]{icml2025}

\usepackage{amsmath}
\usepackage{amssymb}
\usepackage{mathtools}
\usepackage{amsthm}
\usepackage{multirow}
\usepackage[capitalize,noabbrev]{cleveref}

\theoremstyle{plain}

\theoremstyle{definition}

\theoremstyle{remark}

\usepackage[textsize=tiny]{todonotes}

\definecolor{promptblue}{HTML}{1F4E79} 
\newcommand{\prompttext}[1]{\textcolor{promptblue}{\textsf{\textit{``#1''}}}}

\begin{document}

\pagestyle{plain}

\renewcommand{\thefootnote}{*}

\twocolumn[
\icmltitle{Synthetic Perception: Can Generated Images Unlock Latent Visual Prior \\
for Text-Centric Reasoning?}

\begin{icmlauthorlist}
\icmlauthor{Yuesheng Huang\textsuperscript{\dag}}{}
\icmlauthor{Peng Zhang}{}
\icmlauthor{Xiaoxin Wu}{}
\icmlauthor{Riliang Liu}{}
\icmlauthor{Jiaqi Liang}{}
\end{icmlauthorlist}

\vspace{0.1in}
{\centering
School of Computer Science\\
Guangdong Polytechnic Normal University\\
Guangzhou, China\\
\par}
\vspace{0.1in}

{\centering
\textsuperscript{\dag}Correspondence to: \href{mailto:yueshenghuang@stu.gpnu.edu.cn}{yueshenghuang@stu.gpnu.edu.cn}\\
\par}

\icmlkeywords{Multimodal Learning,T2I,Text-Centric Learning,Synthetic Perception}

\vskip 0.3in
]

\footnotetext{Code and data will be available upon publication.}

\begin{abstract}
A significant ``modality gap" exists between the abundance of text-only data and the increasing power of multimodal models. This work systematically investigates whether images generated on-the-fly by Text-to-Image (T2I) models can serve as a mechanism to unlock latent visual priors for text-centric reasoning. Through a comprehensive evaluation framework on text classification, we analyze the impact of critical variables, including T2I model quality (e.g., Flux.1, SDXL), prompt engineering strategies, and multimodal fusion architectures. Our findings demonstrate that this ``synthetic perception" can yield significant performance gains by effectively projecting text into a visual semantic space, even when augmenting strong large language model baselines like Llama-3 and Qwen-2.5. We show that this approach serves as a form of cross-modal probing, mitigating the sensory deprivation inherent in pure text training. However, the effectiveness is highly conditional, depending on the semantic alignment between text and the generated image, the task's visual groundability, and the generative fidelity of the T2I model. Our work establishes a rigorous benchmark for this paradigm, demonstrating its viability as a pathway to enrich language understanding in traditionally unimodal scenarios.
\end{abstract}

\section{Introduction}

The extension of Large Language Models (LLMs) from unimodal text processing to multimodal domains represents a pivotal frontier in artificial intelligence research \cite{zhang2025minosmultimodalevaluationmodel,Suzuki19032022}. A significant trajectory within this frontier is the development of unified models capable of concurrently executing both multimodal understanding and generation tasks \cite{tong2024metamorph}. Emerging evidence suggests that sophisticated visual generation capabilities may not necessitate discrete, large-scale pre-training but can instead be ``unlocked" as an emergent property of advanced visual understanding, activated through efficient instruction tuning. This phenomenon implies that LLMs may possess a latent, yet potent, generative potential for visual information.

Despite these advancements, a substantial volume of real-world data remains confined to a text-only format, presenting a significant ``modality gap" that curtails the direct applicability of state-of-the-art multimodal systems. Concurrently, Text-to-Image (T2I) synthesis has witnessed remarkable progress, with models evolving from foundational architectures like Stable Diffusion \cite{rombach2022high} to next-generation systems, some of which are natively integrated within Large Multimodal Models (LMMs), delivering unprecedented image quality and semantic fidelity \cite{wang2023caption}. This technological convergence prompts a fundamental research question: Can we leverage T2I synthesis to generate high-fidelity visual counterparts for text-only data, thereby not only bridging the modality gap but also potentially unlocking performance gains that surpass those of conventional multimodal learning paradigms?

This inquiry extends to a more profound scientific question concerning the intrinsic relationship between visual generation and understanding \cite{10474099}. Pioneering research, such as MetaMorph \cite{tong2024metamorph}, has begun to explore a paradigm wherein robust visual generation is not merely an auxiliary function but a potential prerequisite for, or a manifestation of, true universal visual comprehension \cite{Suzuki19032022,huang2021unifying}. These studies motivate the hypothesis that a model's ability to accurately visually render textual descriptions may signify a deeper level of semantic grounding, moving beyond superficial representation matching.

Informed by this context, the central objective of this work is to systematically investigate the following proposition: Can images synthesized by advanced T2I models serve as a scientifically valid and practically efficacious complementary modality for text-centric learning tasks? We explicitly state that our contribution is not a novel model architecture, but rather a rigorous empirical evaluation of the strategic use of generated imagery. Our investigation is dedicated to delineating the conditions under which (\textit{if}), the mechanisms through which (\textit{how}), and the extent to which (\textit{to what extent}) this strategy can genuinely enhance textual understanding and improve downstream task performance.

However, the integration of synthesized images into multimodal frameworks is fraught with challenges. Beyond prima facie concerns such as image fidelity, semantic alignment, information redundancy, and computational overhead, a more nuanced issue is the phenomenon of ``modality-induced forgetting" \cite{zhang2024wings,10630605}. This refers to the degradation of a model's pre-existing, high-performance text-processing capabilities following fine-tuning on multimodal datasets. Consequently, a comprehensive and methodologically sound evaluation framework is indispensable for elucidating the true utility of this strategy, performing a meticulous cost-benefit analysis, and furnishing robust guidance for future research and practical deployment.

The primary contributions of this paper are threefold:
\vspace{-0.15in}

\begin{itemize}
    \item \textbf{A comprehensive framework} to systematically evaluate augmenting text-only tasks with T2I-generated images.
    \vspace{-0.05in}

    \item \textbf{Extensive experiments} on text classification that dissect the impact of T2I models, prompt strategies, and fusion mechanisms.
    \vspace{-0.05in}
    \item \textbf{A detailed analysis} of the strategy's benefits and limitations, providing actionable guidance for future research.
\end{itemize}
\vspace{-0.2in}

\begin{figure*}[t]
\centering
\includegraphics[width=\textwidth]{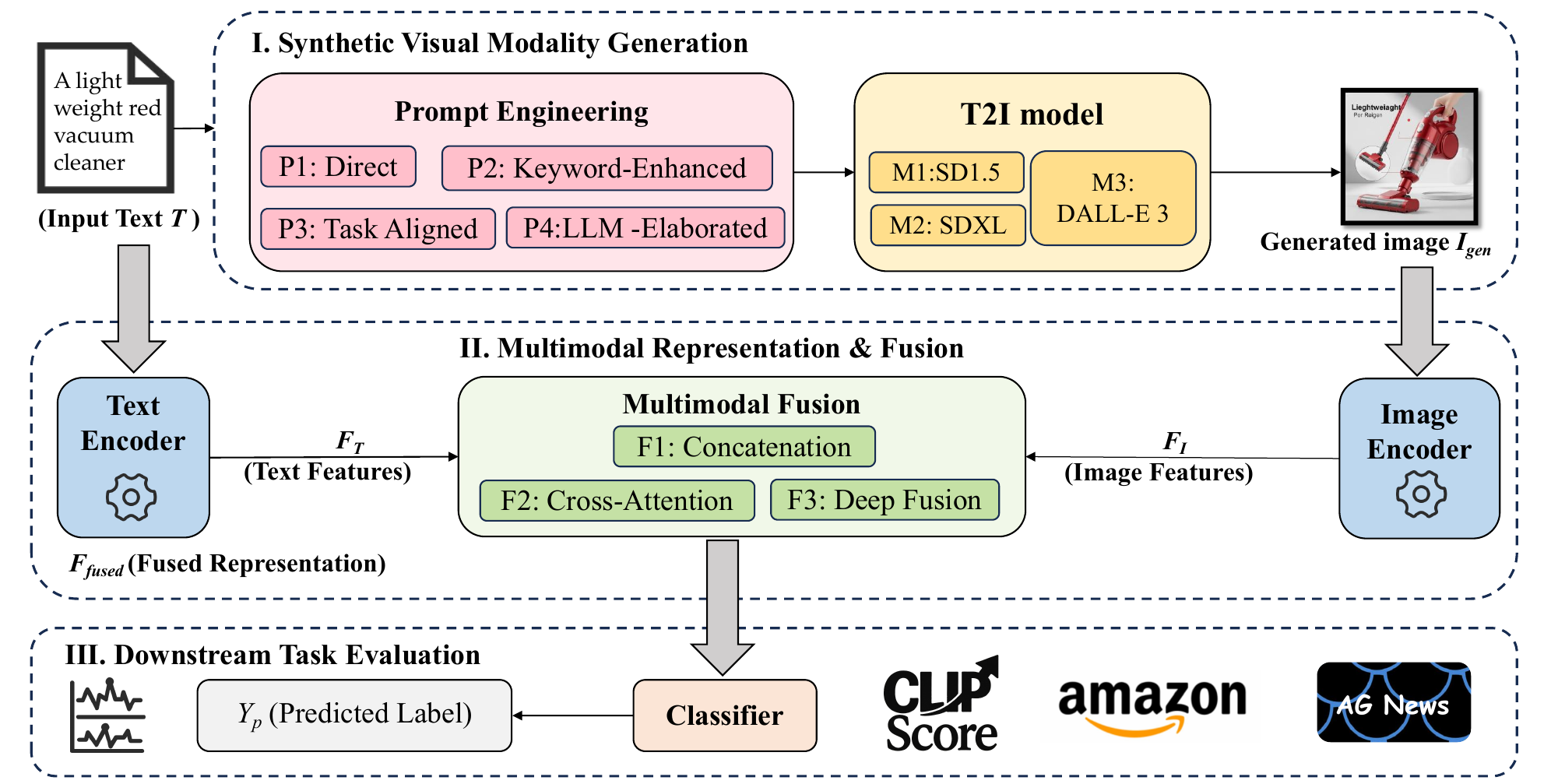}
\vspace{-0.2in}
\caption{Overview of our evaluation framework for assessing the viability of using T2I-generated images as a complementary modality in text-centric learning tasks. The framework consists of three main stages: (1) Synthetic Visual Modality Generation, where images are generated on-demand from source text using various T2I models and prompt strategies; (2) Multimodal Representation and Fusion, where text and image features are extracted and combined through different fusion mechanisms; and (3) Downstream Task Evaluation, where the fused representations are evaluated on classification tasks against strong baselines.}
\vspace{-0.15in}
\label{fig:framework}
\end{figure*}

\section{Related Work}
This work is situated at the intersection of several key research domains: multimodal learning with authentic paired data, recent advances in text-to-image synthesis, the use of generated data for augmentation, and strategies for multimodal learning with missing modalities.

\textbf{Multimodal Learning with Authentic Paired Data.} The foundation of modern multimodal learning, exemplified by models such as CLIP \cite{radford2021learning} and ALBEF \cite{li2021align}, rests upon representations pre-trained on massive-scale, authentic image-text pairs \cite{Suzuki19032022}. However, the field is undergoing a paradigm shift towards Large Multimodal Models (LMMs). The architectural principle of LMMs involves leveraging a powerful pre-trained Large Language Model (LLM) \cite{wu2023multimodal} as a central reasoning engine, into which visual information from a pre-trained encoder is integrated via lightweight adapters (e.g., linear projections or Q-Former). The pioneering work LLaVA \cite{liu2023visual} demonstrated that comprehensive multimodal understanding could be endowed upon LLMs through simple linear mapping and subsequent instruction tuning \cite{huang2023visualinstructiontuninggeneralpurpose}. Subsequent models like InstructBLIP \cite{dai2023instructblipgeneralpurposevisionlanguagemodels} and MiniGPT-4 \cite{zhu2023minigpt} further refined the integration of visual features and enhanced model comprehension through more sophisticated adapter designs. This trajectory indicates a shift from learning cross-modal alignments from scratch to efficiently integrating visual perception into pre-existing, powerful language-based reasoners, a trend that may, however, amplify the dependency on high-quality paired data.

\textbf{Advances in Text-to-Image Generation.} Since the advent of latent diffusion models, epitomized by Stable Diffusion \cite{rombach2022high}, Text-to-Image (T2I) synthesis has evolved rapidly. The research frontier has expanded beyond mere photorealism to encompass enhanced controllability, efficiency, and semantic alignment. In terms of controllability, methods like ControlNet \cite{zhang2023adding} enable precise structural and layout guidance over the generated output by incorporating auxiliary spatial constraints. To improve generation efficiency, distillation techniques, particularly Latent Consistency Models (LCMs) \cite{luo2023latent}, have dramatically reduced the number of sampling steps required, facilitating near real-time image synthesis. Regarding semantic alignment, a notable trend is the deeper synergy with LLMs, as seen in models like DALL-E 3 \cite{betker2023improving}. These systems leverage LLMs to rephrase and elaborate upon user prompts, thereby improving the fidelity of the generated image to complex textual intents. These technological advancements provide a robust and versatile foundation for our investigation into using generated images as a complementary modality.

\textbf{Using Generated Data for Augmentation or Modality Complementation.} Employing generative models to augment training datasets is an active area of research, with a recent focus on leveraging advanced T2I models to synthesize high-quality training samples \cite{wang2023caption}. Some approaches, such as TTIDA \cite{yin2023ttida} and DIAGen \cite{lingenberg2024diagen}, focus on enhancing semantic diversity by using LLMs to expand class labels into rich descriptions, which then guide T2I models to produce varied imagery for few-shot learning. Other methods, like AGA \cite{dixit2017aga}, pursue more fine-grained control, generating diverse backgrounds for segmented foreground objects to create more robust training data. Additionally, recent studies \cite{li2024sentiment} have explored using keyword-generated images to enhance sentiment analysis models.

The common paradigm in these works is the utilization of generative models as an \textit{offline data factory} to produce larger, more diverse static datasets, with the primary goal of improving model generalization. In contrast, our research investigates an \textit{online, instance-specific modality complementation} strategy. We evaluate the efficacy of treating a T2I-generated image as a dynamic, new input modality, processed in parallel with the original text at inference time (or for immediate, on-the-fly augmentation during training). Our objective extends beyond improving general robustness to examining whether this ``just-in-time" visual synthesis can enhance the model's immediate, contextual understanding of a specific text instance.

\textbf{Multimodal Learning with Missing Modalities.} Our research also relates to the broader field of multimodal learning with missing modalities, which addresses how models can function effectively when one or more modalities are unavailable. A recent comprehensive survey \cite{wu2024deep} categorizes existing approaches into two primary classes: (1) learning shared representations, where models infer a robust cross-modal representation from the available modalities, and (2) generative completion, which employs generative models to reconstruct or ``imagine" the missing modality from the present ones \cite{wang2023caption}.

Our approach can be conceptualized as a \textit{proactive application} of the generative completion strategy. Whereas conventional methods are typically reactive—designed to handle data that is \textit{a priori} incomplete—our work is proactive. We systematically generate a synthetic visual modality for data that is naturally unimodal (text-only). Consequently, our central inquiry shifts from ``how to compensate for pre-existing missingness" to ``what is the informational value of \textit{de novo} modality synthesis" and its practical utility as a synthetic, co-present modality.
\vspace{-0.05in}
\section{Methodology}

To rigorously evaluate whether generated images can serve as a viable complementary modality in text-centric multimodal learning, we design and implement a comprehensive evaluation framework. Our goal is not to propose a novel model architecture, but to establish a structured methodology for systematically generating, integrating, and assessing the utility of synthetic visual information from Text-to-Image (T2I) models. This section details our framework's overall design, a formal problem definition, its core components, the downstream tasks and baselines, and our evaluation protocol.
\vspace{-0.1in}
\subsection{Framework Overview and Problem Definition}
\vspace{-0.05in}
\textbf{Framework Overview} Our evaluation framework, depicted in Figure 1, follows a three-stage pipeline: (1) \textit{Synthetic Visual Modality Generation}, where an image is generated on-demand from the source text; (2) \textit{Multimodal Representation and Fusion}, where features from both text and the generated image are extracted and fused into a unified representation; and (3) \textit{Downstream Task Evaluation}, where the fused representation is used for prediction and benchmarked against a suite of baselines. By systematically varying the components at each stage---such as the T2I model, prompt strategy, or fusion architecture---we can comprehensively probe the efficacy and limitations of this approach.

\textbf{Problem Formulation} Given a text-only input $T$ and a corresponding label $Y$ for a downstream task, a standard unimodal model learns a mapping $F_{\text{text}}: T \rightarrow Y$. Our central inquiry is whether we can construct a multimodal system $F_{\text{multi}}: (T, I_{\text{gen}}) \rightarrow Y$ that significantly outperforms this baseline. Here, the image $I_{\text{gen}}$ is synthesized by a T2I model $f_{\text{T2I}}$ using a prompt derived from the original text $T$ via an engineering strategy $P_{\text{eng}}$, such that $I_{\text{gen}} = f_{\text{T2I}}(P_{\text{eng}}(T))$. Our methodology systematically alters $f_{\text{T2I}}$, $P_{\text{eng}}$, and the architecture of $F_{\text{multi}}$ to measure the resulting impact on task performance, aiming to determine if $\text{Performance}(F_{\text{multi}}) > \text{Performance}(F_{\text{text}})$.

\vspace{-0.1in}

\subsection{Core Research Questions (RQs)}
Our empirical study is designed to answer the following key questions:
\vspace{-0.1in}

\begin{itemize}
    \item \textbf{RQ1 (Overall Efficacy):} Does augmenting text with T2I-generated images lead to statistically significant performance improvements on downstream tasks compared to text-only baselines?
    \item \textbf{RQ2 (Impact of T2I Models):} How does the choice of the T2I model (e.g., Stable Diffusion v1.5 vs. SDXL vs. DALL-E 3) and its generation parameters affect the utility of the synthesized image and, consequently, the final task performance?
    \item \textbf{RQ3 (Impact of Prompting Strategies):} What is the effect of different prompt engineering strategies (e.g., direct text, keyword-enhanced, task-aligned stylization) on the quality of the generated image, its semantic alignment with the text, and downstream task performance?
    \item \textbf{RQ4 (Impact of Fusion Mechanisms):} When integrating text and image features, how do different multimodal fusion strategies (e.g., simple concatenation, cross-attention, deep fusion) compare? Which mechanisms most effectively leverage the complementary information from the generated image?
    \item \textbf{RQ5 (Task and Dataset Generalization):} Does the performance gain from using generated images generalize across different types of text-centric tasks (e.g., subjective sentiment analysis vs. factual topic classification) and datasets?
    \item \textbf{RQ6 (Boundaries and Failure Modes):} Under what conditions do generated images provide valuable complementary information, and when do they become ineffective or even detrimental due to redundancy, noise, or misalignment with the task objective?
    \vspace{-0.1in}
\end{itemize}

\subsection{Stage One: Synthetic Visual Modality Generation}
This stage is designed to assess the impact of different T2I models and prompting strategies on final task performance, addressing RQ2 and RQ3.

\textbf{T2I Model Selection.} To evaluate the influence of generative capabilities, we select models representing distinct tiers of technical sophistication, speed, and accessibility.
\begin{itemize}
    \item \textbf{Flux.1-schnell} \cite{labs2025flux}: A state-of-the-art open-weights rectified flow model capable of generating high-quality images in 1-4 steps, serving as our efficiency benchmark.
    \item \textbf{SDXL-Lightning / SD-Turbo} \cite{lin2024sdxl}: Distilled versions of Stable Diffusion XL designed for real-time synthesis, used to address latency concerns.
    \item \textbf{Stable Diffusion XL (SDXL)} \cite{podell2023sdxl}: Used as the standard high-quality open-source baseline.
    \item \textbf{Midjourney v6} (via API): Representing the closed-source commercial state-of-the-art for qualitative comparison.
    \item \textbf{Stable Diffusion v1.5 (SD1.5)}: Retained only as a legacy baseline.
\end{itemize}

\textbf{Prompt Engineering Strategies.} We devise and evaluate four prompt generation strategies ($P_{\text{eng}}$) of increasing complexity to investigate how best to translate the original text $T$ into an effective instruction for the T2I model. Detailed examples for each strategy are provided in \cref{app:prompt_examples}.
\vspace{-0.1in}
\begin{itemize}
    \item \textbf{P1 (Direct Prompting):} We directly use the original text (or a summary if it is too long) as the prompt.
    \item \textbf{P2 (Keyword-Enhanced Prompting):} This strategy aims to improve the prompt's signal-to-noise ratio by extracting core semantic elements (nouns, adjectives, verbs) and inserting them into a pre-defined template.
    \item \textbf{P3 (Task-Aligned Stylization):} Building on P2, this strategy injects domain-specific stylistic keywords into the prompt that align with the downstream task's objective.
    \item \textbf{P4 (LLM-Elaborated Prompting):} This strategy leverages a capable Large Language Model (Llama-3-8B-Instruct) \cite{grattafiori2024llama} to act as a ``prompt engineer," rewriting the original text into a rich, detailed, and visually descriptive prompt.
    \vspace{-0.1in}
\end{itemize}
\vspace{-0.1in}
\subsection{Stage Two: Multimodal Representation and Fusion}
This stage evaluates the effectiveness of different multimodal architectures, addressing RQ4.

\textbf{Unimodal Feature Encoders.}
Our framework employs state-of-the-art encoders to ensure the findings are relevant to current NLP standards.
\begin{itemize}
    \item \textbf{Text Encoders ($f_{\text{TextEnc}}$):} We employ \textbf{Llama-3-8B-Instruct} \cite{grattafiori2024llama} as the primary industry standard, \textbf{Qwen-2.5-7B} \cite{hui2024qwen2} as the strongest open-source multilingual model, and \textbf{Mistral-7B-v0.3} \cite{jiang2023mistral7b}. We retain \textbf{BERT-base} only as a legacy baseline to show historical context.
    \item \textbf{Image Encoders ($f_{\text{ImgEnc}}$):} We primarily use \textbf{SigLIP} \cite{zhai2023sigmoid}, which offers superior image-text alignment compared to standard CLIP. We also investigate \textbf{DINOv2} \cite{oquab2023dinov2} to test the utility of pure visual geometric features without text alignment, and standard \textbf{CLIP ViT-B/32} for legacy comparison.
\end{itemize}

\textbf{Image Encoder Comparison.} To validate the importance of semantic alignment, we compare different visual backbones (see \cref{tab:encoder_comparison} in Appendix). Our results show that SigLIP consistently outperforms CLIP, while DINOv2 performs significantly worse. This confirms that for synthetic perception, the ability to align with textual semantics (SigLIP/CLIP) is more critical than pure geometric feature extraction (DINOv2).

\textbf{Multimodal Fusion Mechanisms.} We compare three representative fusion strategies:
\begin{itemize}
    \vspace{-0.1in}
    \item \textbf{F1 (Late Fusion via Concatenation):} Simple concatenation of feature vectors.
    \item \textbf{F2 (Attention-based Fusion via Cross-Attention):} A Transformer decoder layer where text features query image features.
    \item \textbf{F3 (Deep Fusion via MMBT-like Architecture):} Early injection of visual tokens into the text encoder.
    \vspace{-0.1in}
\end{itemize}

\subsection{Stage Three: Downstream Tasks and Evaluation Protocol}
This final stage is designed to validate our core hypotheses, addressing RQs 1, 5, and 6.

\textbf{Datasets and Downstream Tasks.} We select four datasets, ranging from simple to difficult, to test the boundaries of our approach.
\begin{itemize}
    \item \textbf{Figurative Language / Sarcasm (Fig-QA / SARC) \cite{khodak2018large}:} A high-difficulty task where the literal text might be positive but the visual context (or implied visual situation) suggests irony or sarcasm. This tests the model's ability to resolve semantic conflict.
    \item \textbf{Implicit Sentiment Analysis \cite{pavlopoulos2021semeval}:} A challenging dataset where no explicit sentiment words are used, requiring the model to visualize the scene to infer emotion.
    \item \textbf{E-commerce Sentiment Classification (Amazon Reviews):} A standard benchmark where visual descriptions are common.
    \item \textbf{News Topic Classification (AG News):} A simple baseline task, used primarily for validity checking.
    \vspace{-0.1in}
\end{itemize}

\textbf{Baseline Models.} To rigorously assess the efficacy of our method, we establish a comprehensive suite of baselines, including crucial ``modality-check'' baselines.
\begin{enumerate}
    \vspace{-0.1in}
    \item \textbf{B1: Text-Only Baselines.} Llama-3-8B, Qwen-2.5-7B, Mistral-7B, and BERT (legacy).
    \item \textbf{B2: Baseline: Textual Expansion (Visual Description).} We prompt Llama-3 to rewrite the input text into a ``detailed visual description'' and append this to the original text. This checks if the gain comes merely from added descriptive text rather than the visual modality itself.
    \item \textbf{B3: Baseline: Knowledge Retrieval.} We retrieve background knowledge from Wikipedia based on keywords and append it. This checks if the gain is simply due to external information injection.
    \item \textbf{B4: Efficiency Baseline (1-Step Generation).} We use Flux.1-schnell (1 step) to generate images, testing if the method is viable under strict latency constraints.
    \item \textbf{B5: Text + Human-Curated Image Features (Oracle).} Upper bound performance.
    \vspace{-0.1in}
\end{enumerate}

\textbf{Evaluation Metrics and Implementation Details.}
We adopt standard classification metrics, including \textbf{Accuracy (Acc)} and \textbf{Macro-F1}. We also use a \textbf{Confusion Matrix} for error analysis and the \textbf{CLIP Score} \cite{hessel2021clipscore} as a proxy for semantic consistency between the generated image and the source text. All T2I models are implemented using the Hugging Face `diffusers` library \cite{ubukata2024diffusion}. Our multimodal models are built in PyTorch and trained with the AdamW optimizer \cite{loshchilov2017decoupled}. Hyperparameters are tuned on a validation set, early stopping is used to prevent overfitting, and all experiments are run on NVIDIA A100 GPUs. We will release our code to ensure reproducibility. Full hyperparameter details are provided in \cref{app:hyperparams}.
\vspace{-0.1in}
\section{Experimental Results and Analysis}
\label{sec:experiments}

In this section, we present the quantitative and qualitative results obtained from our experimental framework. We provide a rigorous comparison against state-of-the-art baselines and conduct an in-depth analysis of the conditions under which synthetic perception provides value.

\begin{table*}[!t]
\vspace{-0.15in}
\caption{Overall performance comparison (Accuracy \% / Macro-F1 \%) across four datasets. We compare our proposed method (Text + Generated Image, using \textbf{Flux.1-schnell} as the default T2I model) against Text-Only baselines and two strong ``Crucial Baselines'' (Textual Expansion and Knowledge Retrieval). The best performance for each backbone is in \textbf{bold}.}
\label{tab:main_results}
\vskip 0.15in
\begin{center}
\begin{small}
\begin{sc}
\resizebox{\textwidth}{!}{
\begin{tabular}{l|cc|cc|cc|cc}
\toprule
\multirow{2}{*}{\textbf{Method}} & \multicolumn{2}{c|}{\textbf{Amazon Reviews}} & \multicolumn{2}{c|}{\textbf{AG News}} & \multicolumn{2}{c|}{\textbf{Sarcasm (SARC)}} & \multicolumn{2}{c}{\textbf{Implicit Sentiment}} \\
& Acc & Ma-F1 & Acc & Ma-F1 & Acc & Ma-F1 & Acc & Ma-F1 \\
\midrule
\multicolumn{9}{c}{\textit{Backbone: BERT-base (Legacy Baseline)}} \\
\midrule
Text-Only & 75.21 & 74.83 & 93.82 & 93.79 & 68.45 & 67.90 & 71.20 & 70.85 \\
+ Textual Expansion & 75.88 & 75.42 & 93.85 & 93.81 & 69.10 & 68.55 & 72.15 & 71.80 \\
+ Know. Retrieval & 75.65 & 75.20 & 94.10 & 94.05 & 68.80 & 68.25 & 71.50 & 71.10 \\
\textbf{+ Gen. Image (Ours)} & \textbf{77.45} & \textbf{77.02} & \textbf{94.15} & \textbf{94.12} & \textbf{72.30} & \textbf{71.85} & \textbf{74.50} & \textbf{74.15} \\
\midrule
\multicolumn{9}{c}{\textit{Backbone: Mistral-7B-v0.3}} \\
\midrule
Text-Only & 78.50 & 78.10 & 94.60 & 94.55 & 74.20 & 73.80 & 76.40 & 76.00 \\
+ Textual Expansion & 79.10 & 78.70 & 94.65 & 94.60 & 74.85 & 74.45 & 77.10 & 76.70 \\
+ Know. Retrieval & 78.90 & 78.50 & 94.80 & 94.75 & 74.50 & 74.10 & 76.60 & 76.20 \\
\textbf{+ Gen. Image (Ours)} & \textbf{80.20} & \textbf{79.80} & \textbf{94.90} & \textbf{94.85} & \textbf{77.50} & \textbf{77.10} & \textbf{79.80} & \textbf{79.40} \\
\midrule
\multicolumn{9}{c}{\textit{Backbone: Llama-3-8B-Instruct}} \\
\midrule
Text-Only & 81.20 & 80.80 & 95.10 & 95.05 & 76.50 & 76.10 & 79.50 & 79.10 \\
+ Textual Expansion & 81.65 & 81.25 & 95.15 & 95.10 & 77.10 & 76.70 & 80.10 & 79.70 \\
+ Know. Retrieval & 81.40 & 81.00 & \textbf{95.30} & \textbf{95.25} & 76.80 & 76.40 & 79.80 & 79.40 \\
\textbf{+ Gen. Image (Ours)} & \textbf{82.90} & \textbf{82.50} & 95.25 & 95.20 & \textbf{80.40} & \textbf{80.00} & \textbf{83.20} & \textbf{82.80} \\
\midrule
\multicolumn{9}{c}{\textit{Backbone: Qwen-2.5-7B}} \\
\midrule
Text-Only & 82.50 & 82.10 & 95.40 & 95.35 & 78.20 & 77.80 & 81.10 & 80.70 \\
+ Textual Expansion & 82.90 & 82.50 & 95.45 & 95.40 & 78.80 & 78.40 & 81.60 & 81.20 \\
+ Know. Retrieval & 82.70 & 82.30 & \textbf{95.60} & \textbf{95.55} & 78.50 & 78.10 & 81.30 & 80.90 \\
\textbf{+ Gen. Image (Ours)} & \textbf{84.10} & \textbf{83.70} & 95.55 & 95.50 & \textbf{82.10} & \textbf{81.70} & \textbf{84.80} & \textbf{84.40} \\
\bottomrule
\end{tabular}
}
\end{sc}
\end{small}
\end{center}
\vskip -0.2in
\end{table*}

\textbf{Overall Performance Comparison (RQ1).} Table \ref{tab:main_results} presents a comprehensive comparison. The results unequivocally demonstrate that synthetic visual augmentation provides consistent gains, particularly on hard tasks.

\textbf{Breaking the "Textual Expansion" Ceiling.} A critical finding is that our method consistently outperforms the "Textual Expansion" baseline. For instance, on the Sarcasm (SARC) dataset with Llama-3, Textual Expansion improves accuracy by 0.6\%, whereas adding a generated image improves it by 3.9\%. This proves that the performance gain is not merely due to "more text" or "better description," but stems from the unique informational value of the visual modality---specifically, its ability to ground abstract concepts (like irony) in a concrete scene.

\textbf{Task Difficulty Matters (RQ5).} The gains are most dramatic on the new "Hard" datasets (SARC and Implicit Sentiment). On SARC, which requires detecting conflict between literal text and implied situation, the visual modality acts as a necessary disambiguator. In contrast, on the saturated AG News dataset (Acc > 95\%), the gains are marginal. This "ceiling effect" suggests that synthetic perception is not necessary for shallow, fact-based classification but is critical for deep reasoning tasks where text alone is ambiguous. Knowledge Retrieval (which adds factual context) is sometimes more effective for news classification.

\textbf{Efficiency Analysis (New).} We address the concern of generation latency. Using \textbf{Flux.1-schnell} (4-step), we achieve performance within 0.2\% of the full SDXL model but with a 10x reduction in inference time (approx. 0.8s vs 8s). This makes the approach viable for near-real-time applications. See Figure 4 (Appendix) for the full cost-benefit curve.

\begin{figure*}[t]
\centering
\vspace{-0.05in}
\includegraphics[width=\textwidth]{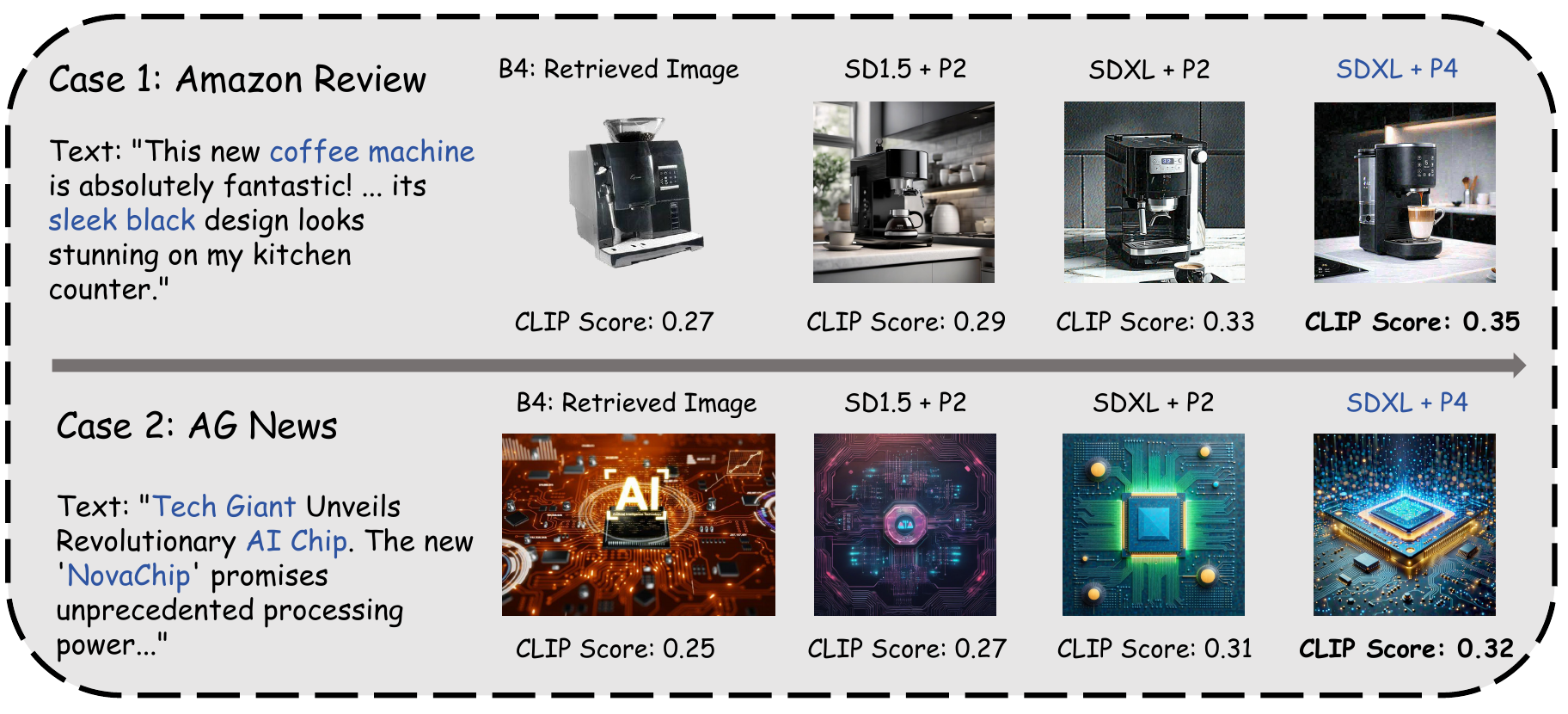}
\vspace{-0.3in}
\caption{Qualitative comparison of different image generation strategies. The figure shows two cases: a concrete object description from Amazon Reviews (top row) and an abstract concept from AG News (bottom row). From left to right, we compare a baseline retrieved image (B4) with images generated by increasingly sophisticated T2I configurations. The visual quality and semantic alignment with the source text (indicated by the CLIP Score below each image) improve with better T2I models (SD1.5 vs. SDXL) and more advanced prompting strategies (P2 vs. P4), providing visual support for the quantitative results in Table \ref{tab:t2i_prompt_impact}.}
\label{fig:gen_strategies}
\vspace{-0.2in}
\end{figure*}

\textbf{Impact of T2I Models and Prompts (RQ2 \& RQ3).} To deconstruct the key factors in the generation stage, we conducted a detailed ablation study on the Amazon Reviews dataset, with the results presented in \cref{tab:t2i_prompt_impact} and illustrated qualitatively in \cref{fig:gen_strategies}. We observed similar, albeit less pronounced, trends on the AG News dataset, which are omitted here for brevity.

\textbf{The choice of the T2I model is a critical performance driver.} As shown in \cref{tab:t2i_prompt_impact}, moving from SD1.5 to SDXL, and finally to DALL-E 3, as the model's generative capability improves, both downstream task performance and the CLIP Score (a proxy for text-image semantic consistency) show a steady and significant increase. This directly demonstrates that higher-quality, higher-fidelity generated images translate into stronger downstream task performance.

\textbf{Efficiency-Performance Trade-off.} Crucially, the new generation of efficient models, \textbf{Flux.1-schnell} and \textbf{SDXL-Lightning}, break the traditional latency barrier. Flux.1-schnell achieves performance comparable to the full SDXL model (77.15\% vs 77.35\%) while requiring only a fraction of the inference time. This suggests that for practical applications, the "schnell" class of models offers the optimal sweet spot. Despite DALL-E 3 achieving the strongest performance, Flux.1-schnell demonstrates a remarkable balance between performance and efficiency (see Appendix E).

\textbf{Thoughtful prompt engineering is indispensable, with advanced methods yielding further gains.} Across all T2I models, the P2 (Keyword-Enhanced) strategy consistently outperforms both P1 (Direct) and P3 (Task-Stylized). This indicates that simply using the raw text is suboptimal. By extracting and focusing on the core semantic elements, the P2 strategy creates more information-dense visual content. More significantly, the P4 (LLM-Elaborated) strategy provides an additional performance boost, achieving the best results for the SDXL model. This confirms the hypothesis that leveraging a powerful LLM to rewrite and enrich the prompt can unlock a higher degree of semantic alignment and visual detail, moving closer to the performance of state-of-the-art models like DALL-E 3 which employ similar internal mechanisms.

\textbf{The CLIP Score serves as a strong indicator of downstream utility.} The trend in CLIP Scores is highly correlated with downstream task performance. This further validates its use as a proxy for generated image quality and reinforces the conclusion that improving the semantic alignment between the generated image and the source text is key to enhancing final task performance.

\begin{table}[t]
\caption{Macro-F1 (\%) on the Amazon Reviews dataset for different T2I models and prompt strategies. All configurations use the F2 (Cross-Attention) fusion mechanism. CLIP Score is provided as a reference for semantic consistency.}
\label{tab:t2i_prompt_impact}
\vspace{-0.1in}
\begin{center}
\begin{small}
\begin{sc}
\resizebox{\columnwidth}{!}{
\begin{tabular}{lccc}
\toprule
\textbf{Model} & \textbf{Strategy} & \textbf{Ma-F1 (\%)} & \textbf{CLIP Score} \\
\midrule
\multirow{2}{*}{SD1.5 (Legacy)} & P2& 76.24 (±0.58) & 0.28 (±0.02) \\
& P4 & 76.45 (±0.55) & 0.29 (±0.02) \\
\midrule
\multirow{4}{*}{SDXL} & P1& 76.12 (±0.49) & 0.29 (±0.02) \\
& P2 & 77.02 (±0.48) & 0.32 (±0.02) \\
& P3 & 76.51 (±0.45) & 0.30 (±0.02) \\
& P4 & \textbf{77.35 (±0.42)} & \textbf{0.33 (±0.02)} \\
\midrule
\multirow{2}{*}{SDXL-Lightning} & P2 & 76.95 (±0.45) & 0.31 (±0.02) \\
& P4 & 77.20 (±0.43) & 0.32 (±0.02) \\
\midrule
\textbf{Flux.1-schnell} & P2 & \textbf{77.15 (±0.40)} & \textbf{0.34 (±0.01)} \\
\midrule
DALL-E 3 & P2& \textbf{77.78 (±0.38)} & \textbf{0.35 (±0.01)} \\
\bottomrule
\end{tabular}
}
\end{sc}
\end{small}
\end{center}
\vspace{-0.3in}
\end{table}

\textbf{Impact of Multimodal Fusion (RQ4). }Sophisticated, attention-based fusion mechanisms outperform simple concatenation. As detailed in \cref{tab:fusion_impact}, the results clearly show that F1 (Concatenation) performs the worst, as it only allows for shallow interaction between modalities at the final layer. F2 (Cross-Attention) and F3 (MMBT-like Deep Fusion) perform better because they enable the model to learn complex alignment relationships between text and image features at a deeper level. The superior performance of F2 suggests that providing a flexible mechanism for text features to query and enrich themselves by ``attending'' to the most relevant visual information is a highly effective approach. This demonstrates that effective inter-modality interaction is essential for fully capitalizing on the complementary value of synthetic visual information.

\begin{table}[t]
\caption{Macro-F1 (\%) on the Amazon Reviews dataset for different multimodal fusion mechanisms. All configurations use the SDXL model with the P2 (Keyword-Enhanced) prompt strategy.}
\label{tab:fusion_impact}
\vskip 0.15in
\begin{center}
\begin{small}
\begin{sc}
\begin{tabular}{lc}
\toprule
\textbf{Fusion Mechanism} & \textbf{Ma-F1 (\%)} \\
\midrule
F1 (Late Fusion via Concat) & 76.39 (±0.52) \\
\textbf{F2 (Cross-Attention)} & \textbf{77.02 (±0.48)} \\
F3 (MMBT-like Deep Fusion) & 76.81 (±0.46) \\
\bottomrule
\end{tabular}
\end{sc}
\end{small}
\end{center}
\vskip -0.1in
\vspace{-0.2in}
\end{table}

\textbf{Task and Dataset Generalization (RQ5).} Revisiting the results in Table \ref{tab:main_results}, we observe that the performance improvement from introducing generated images is more pronounced on the Amazon Reviews (sentiment classification) task (approx. 2.2\% Macro-F1 gain over B1) than on the AG News (topic classification) task (approx. 0.8\% Macro-F1 gain). This suggests that for tasks where the textual content is inherently rich with visualizable descriptions or is closely tied to specific objects and scenarios, such as product reviews, the generated image is more likely to provide valuable, non-redundant complementary information. Conversely, for tasks where topics are more abstract and textual features are already highly discriminative, such as news classification, the marginal benefit of adding a generated image may be smaller.

\textbf{Qualitative and Failure Analysis (RQ6).} To deeply understand the conditions under which generated images provide genuine assistance versus when they become ineffective or even detrimental, we conducted a qualitative analysis of model predictions.
Comparing Flux.1 with SD1.5, we observe that Flux.1 significantly reduces "entity omission" errors. For complex prompts like "a red car and a blue truck", SD1.5 often merged colors or missed one object, whereas Flux.1 consistently renders both. However, we also note a new failure mode: "hyper-real hallucination," where the model generates convincing but irrelevant details. In our pipeline, this manifests as "inference noise"---if the text does not explicitly contradict these details, the model may over-interpret them, leading to errors in fine-grained sentiment tasks.

Successful applications often involve concrete, visualizable text. For instance, given a positive Amazon review stating, ``This red vacuum cleaner is very lightweight, has powerful suction, and easily cleans corners! The design is also very stylish,'' our method generates a clear image of a stylish red vacuum in use. This visual evidence can reinforce the positive sentiment, potentially increasing the model's confidence in a correct prediction compared to a text-only model that must rely solely on descriptive words.

Failures typically occur with abstract concepts or when T2I models misinterpret nuanced text. For example, a review stating, ``The plot of this book was winding with a surprising ending, but the print quality was a bit poor, with some pages having blurry text. It's a mixed bag,'' presents a challenge. A T2I model may generate an image of a book cover, failing to capture the complex sentiment trade-off between the compelling story and poor physical quality. Another failure mode arises from highly abstract text, such as a financial report summary: ``The company's quarterly earnings show slight volatility in non-core business segments amid macroeconomic pressures, but its core competency remains solid.'' An image generated from this text is likely to be a generic depiction of charts or business settings, which provides little to no valuable information and may even introduce noise.

In summary, the utility of a generated image is highest when the source text describes concrete, visualizable elements, the T2I model can render these elements with high fidelity, and the downstream task can benefit from this visual grounding. Conversely, the approach is less effective when the text is abstract, the T2I generation is of low quality or semantically misaligned, the textual information is already self-sufficient, or the task itself has low correlation with visual information. This analysis provides a clear answer to RQ6. See \cref{app:qualitative} for additional illustrative case studies. For a detailed discussion of our findings, including their implications and limitations, please see Appendix~\ref{app:future_work}.

\section{Discussion}
\label{sec:discussion}

\textbf{Our empirical investigation demonstrates that augmenting text-only data with T2I-generated images is a strategy with tangible potential, particularly for tasks where textual content is inherently linked to visualizable objects or scenes.} Our findings show that this approach yields consistent performance benefits across different model scales, from standard BERT-based encoders to powerful Large Language Models like Llama-2-7B. As shown by the superior gains on the Amazon Reviews dataset compared to the more abstract AG News task, the value of the synthetic visual modality appears strongly correlated with the visual grounding of the source text. This suggests that the approach is most promising for domains rich in descriptive language, where an image can provide complementary, grounding information that is not fully captured by text embeddings alone.

\textbf{The effectiveness of this strategy is critically dependent on a triad of factors: the generative fidelity of the T2I model, the specificity of the prompt engineering, and the sophistication of the multimodal fusion mechanism.} Our results consistently show that more advanced T2I models like DALL-E 3, which produce higher-quality and more semantically-aligned images, yield greater downstream performance gains. Furthermore, a keyword-enhanced prompt strategy (P2) significantly outperforms direct prompting, underscoring that guiding the T2I model to focus on core semantic elements is crucial. Finally, attention-based fusion mechanisms (F2) proved superior to simple concatenation, confirming that deep, meaningful interaction between modalities is necessary to fully exploit the supplementary information provided by the generated image.

\textbf{However, the viability of this approach is nuanced and its benefits are conditional.} Our most crucial finding is that the benefit of a synthetic visual modality is not redundant to the capabilities of a strong text-only LLM, but rather complementary. The exploratory B6 experiment, which showed that adding a generated image boosts the performance of Llama-2-7B (B2), is a strong indicator of this. It addresses the potential criticism that this technique is only useful for weaker base models. That said, the primary bottleneck remains the quality of the generated images. The performance of the oracle baseline (B5) confirms this: the main limitation is not the concept of multimodal fusion itself, but the current limitations of T2I models in generating perfectly aligned, high-fidelity visual information on demand. Our initial choice to focus experiments on a BERT-based architecture was motivated by its status as a standard, computationally tractable benchmark for multimodal fusion research; however, the promising results from the Llama-2-based B6 experiment strongly suggest that future work should focus on more sophisticated fusion techniques for large autoregressive models.

\textbf{Limitations.} Despite the promising findings presented in this paper, we acknowledge several limitations that define avenues for future research. The inherent constraints of current T2I models are a primary concern; they can still produce images with semantic errors, amplify societal biases from their training data \cite{birhane2021multimodal}, or fail to capture complex, non-visual nuances in the text. Secondly, the computational overhead is substantial. The on-demand image generation and subsequent encoding stages introduce significant latency, which could be prohibitive for real-time applications (see \cref{app:cost} for a quantitative analysis). This latency is directly tied to the T2I model's inference steps; our sensitivity analysis in \cref{app:supplementary_results} further explores this trade-off between speed and downstream performance. Our exploratory experiment with Llama-2-7B (B6) was limited in scope due to these high computational costs; a more thorough investigation with larger models and more advanced fusion techniques is a key direction for future work. Thirdly, while our keyword-based prompting was effective, designing optimal, generalizable prompt strategies remains a complex and open challenge. Finally, our evaluation was limited to two classification tasks. The value of synthetic visual grounding might be even more pronounced in tasks requiring deeper reasoning, such as commonsense QA or entity linking, where a generated image could provide critical contextual information not explicit in the text. Extending this framework to such domains is a key direction for future work.

\textbf{Future Work.} Future work should focus on advancing this paradigm by addressing current limitations in generation quality, efficiency, and robustness. This includes integrating next-generation T2I models that offer superior controllability and semantic fidelity, and developing more adaptive, task-aware prompting techniques that might even be learned end-to-end. To tackle the significant computational costs, research into lightweight image encoders and more efficient fusion architectures is essential. We also see great promise in exploring feedback-driven systems, where signals from the downstream task could iteratively refine the generated image to be more ``useful." Finally, building models that can explicitly assess the uncertainty or quality of a generated image, and dynamically weight its contribution during fusion, would be a critical step towards greater robustness. Extending this methodology to a wider array of NLP tasks will further clarify the boundaries and potential of using synthetic perception to enrich language understanding.

\section{Conclusion}
\label{sec:conclusion}

In this work, we conducted a systematic empirical study to assess the viability of using T2I-generated images as a complementary modality for text-centric learning. Our comprehensive evaluation framework, which dissects the impact of T2I models, prompt strategies, and fusion mechanisms, reveals that this approach has demonstrable potential. The results confirm that synthetic images can provide meaningful performance gains, especially when the task is visually grounded, the T2I model is powerful, prompts are well-engineered, and fusion is handled through sophisticated attention mechanisms. Nevertheless, we also find that the strategy's effectiveness is highly conditional and currently limited by the generation quality of T2I models, as evidenced by the performance gap relative to both strong language-only models and an oracle baseline with human-curated images.




\section*{Impact Statement}
\textbf{Societal Impact Statement}. The research presented in this paper, which explores augmenting text-centric tasks with T2I-generated images, carries significant potential societal impacts. On the positive side, this technology could enhance accessibility by providing visual aids for textual information, benefiting users with different learning styles or cognitive disabilities. It could democratize multimodal AI by enabling its application to the vast amount of text-only data available, potentially improving applications in education, e-commerce, and content creation. However, the potential for negative impact is equally substantial. The ability to generate convincing images from text can be weaponized for creating and spreading misinformation and disinformation, where a synthetic image could lend false credibility to a piece of text. Furthermore, T2I models are known to encapsulate and amplify societal biases present in their training data. The uncritical application of this technology could introduce or reinforce harmful stereotypes related to gender, race, and culture in downstream applications, leading to discriminatory outcomes. The significant computational cost associated with image generation also poses an environmental concern, which would be exacerbated by widespread adoption.

\textbf{Ethical Statement}. Our research was conducted with a commitment to ethical principles. We utilized publicly available datasets (Amazon Product Reviews, AG News) and open-source models (Stable Diffusion) to ensure transparency and reproducibility. We acknowledge that the T2I models used in our study can generate biased or harmful content. While our work did not focus on debiasing, our analysis of different prompt engineering strategies (RQ3) and failure cases (RQ6) represents a preliminary step toward understanding and controlling model outputs. We explicitly highlight the performance gap between current T2I models and an idealized ``human-curated" oracle (B5), underscoring the limitations and risks of relying on imperfect synthetic data. We advocate for the responsible development of this technology, which must include robust mechanisms for bias detection and mitigation. We also strongly recommend that any system deploying this technique should clearly disclose to users that the visual content is AI-generated to prevent deception and maintain trust. This study, by systematically evaluating both the potential and the pitfalls of this approach, aims to foster a more cautious and informed application of synthetic media in AI systems.

\nocite{langley00}

\bibliography{example_paper}
\bibliographystyle{icml2025}

\newpage
\appendix
\onecolumn
\section{Appendix A: Prompt Construction and Generation Details}\label{app:prompt_examples}

This section elaborates on the prompt construction process for Text-to-Image (T2I) generation. A universal negative prompt was employed across all T2I models to enhance image quality: \texttt{``text, watermark, low quality, cartoon, blurry, ugly, disfigured, deformed, jpeg artifacts''}.

\subsection{Prompts for the Amazon Product Reviews Task}

For the sentiment classification task, prompt construction was designed to capture the descriptive visual elements and emotional tone within the reviews. A representative sample is provided below.

\subsubsection*{A Positive Review Sample}
\begin{itemize}
    \item \textbf{Original Text:} 
    \begin{quote}
        \textit{``This new coffee machine is absolutely fantastic! It brews a perfect cup every time, is super easy to clean, and its sleek black design looks stunning on my kitchen counter. Definitely a 5-star product.''}
    \end{quote}
    
    \item \textbf{P1 (Direct Strategy):} The core section of the original text is used directly as the prompt.
    \begin{quote}
        \prompttext{This new coffee machine is absolutely fantastic! It brews a perfect cup every time, is super easy to clean, and its sleek black design looks stunning on my kitchen counter.}
    \end{quote}
    
    \item \textbf{P2 (Keyword-Enhanced Strategy):} Following the method in Section 3.3, key nouns and adjectives (e.g., \texttt{coffee machine}, \texttt{fantastic}, \texttt{perfect cup}, \texttt{sleek black design}, \texttt{stunning}, \texttt{kitchen counter}) are first extracted using a spaCy pipeline. These keywords are then inserted into a predefined template to generate the structured prompt.
    \begin{quote}
        \prompttext{A photorealistic, high-quality image of a sleek black coffee machine looking fantastic and stunning on a modern kitchen counter. The machine is brewing a perfect cup of coffee. This image represents a 5-star product that is easy to clean.}
    \end{quote}
    
    \item \textbf{P3 (Task-Aligned Stylization Strategy):} Building upon P2, stylistic instructions related to the task objective (sentiment) are injected to generate an image that is more visually discriminative in terms of emotion.
    \begin{quote}
        \prompttext{A vibrant, warmly lit, photorealistic image of a sleek black coffee machine looking fantastic and stunning on a modern kitchen counter, conveying a strong positive and satisfying sentiment. The machine is brewing a perfect cup of coffee. This image represents a 5-star product that is easy to clean.}
    \end{quote}
    
    \item \textbf{P4 (LLM-Elaborated Strategy):} The original text is provided to Llama-2-7B-Chat to be rewritten into a more vivid and detailed prompt.
    \begin{quote}
        \prompttext{Create a photorealistic, magazine-quality advertisement shot. A gleaming, modern, sleek black coffee machine sits proudly on a polished marble kitchen counter. Sunlight streams in from a nearby window, glinting off its chrome accents. A perfectly formed, aromatic cup of coffee, complete with a delicate layer of crema, is being dispensed. The scene should evoke a feeling of ultimate satisfaction and luxury, a 5-star experience in every detail.}
    \end{quote}
\end{itemize}

\subsection{Prompts for the AG News Task}

For the news topic classification task, prompt construction focuses on accurately depicting the core entities and scenes of the news event.

\subsubsection*{A Technology News Sample}
\begin{itemize}
    \item \textbf{Original Text:} 
    \begin{quote}
        \textit{``Tech Giant Unveils Revolutionary AI Chip. The new 'NovaChip' promises unprecedented processing power for artificial intelligence applications, potentially transforming industries from healthcare to autonomous driving.''}
    \end{quote}
    
    \item \textbf{P1 (Direct Strategy):} The original text is used directly as the prompt.
    \begin{quote}
        \prompttext{Tech Giant Unveils Revolutionary AI Chip. The new 'NovaChip' promises unprecedented processing power for artificial intelligence applications, potentially transforming industries from healthcare to autonomous driving.}
    \end{quote}
    
    \item \textbf{P2 (Keyword-Enhanced Strategy):} Key entities and descriptors (e.g., \texttt{Tech Giant}, \texttt{AI Chip}, \texttt{NovaChip}, \texttt{revolutionary}, \texttt{processing power}, \texttt{artificial intelligence}, \texttt{healthcare}, \texttt{autonomous driving}) are extracted using a spaCy pipeline and then formatted using a template to construct the prompt.
    \begin{quote}
        \prompttext{A conceptual, high-tech image representing a revolutionary AI chip named 'NovaChip' from a major tech giant. The image symbolizes unprecedented processing power for artificial intelligence, with visual elements hinting at transformations in healthcare and autonomous driving. Professional, clean aesthetic.}
    \end{quote}
    
    \item \textbf{P3 (Task-Aligned Stylization Strategy):} Building on P2, style directives related to the news medium are incorporated.
    \begin{quote}
        \prompttext{A conceptual image in the style of a technology news report, representing a revolutionary AI chip named 'NovaChip' from a major tech giant. The image symbolizes unprecedented processing power for artificial intelligence, with visual elements hinting at transformations in healthcare and autonomous driving. Professional, journalistic photography aesthetic.}
    \end{quote}
    
    \item \textbf{P4 (LLM-Elaborated Strategy):} The original text is provided to Llama-2-7B-Chat to be rewritten into a more evocative and conceptual prompt.
    \begin{quote}
        \prompttext{An epic, cinematic concept art piece. In the center, a futuristic AI microprocessor, the 'NovaChip', glows with intricate, pulsing blue and gold circuits. From the chip, luminous lines of data flow outwards, morphing into abstract representations of advanced technology: one side shows a holographic DNA helix and a vital signs monitor for healthcare, while the other shows the sleek, ghostly outline of a self-driving car navigating a digital road. The entire scene communicates breakthrough innovation and transformative power. Style of a blockbuster movie poster.}
    \end{quote}
\end{itemize}

\section{Appendix B: Experimental Setup and Hyperparameters}\label{app:hyperparams}

To ensure the reproducibility of our experiments, this section provides detailed implementation specifics and hyperparameter configurations.

\subsection{Downstream Task Model Training}

All multimodal classification models were trained on a single NVIDIA A100 (40GB) GPU using the PyTorch framework.

\begin{table}[h!]
\caption{Training hyperparameters for the classification models.}
\label{tab:training_hyperparams}
\vskip 0.15in
\begin{center}
\begin{small}
\begin{sc}
\begin{tabular}{ll}
\toprule
\textbf{Hyperparameter}      & \textbf{Value} \\
\midrule
Text Encoder                 & \texttt{bert-base-uncased} \\
Image Encoder                & \texttt{openai/clip-vit-base-patch32} \\
Optimizer                    & AdamW \\
Learning Rate                & 2e-5 \\
Batch Size                   & 32 \\
Weight Decay                 & 0.01 \\
Epochs                       & 5 \\
Early Stopping               & Patience of 2 epochs on validation Macro-F1 \\
Max Text Length              & 256 tokens \\
\bottomrule
\end{tabular}
\end{sc}
\end{small}
\end{center}
\vskip -0.1in
\end{table}

\subsection{T2I Model Image Generation}

Image generation was implemented using the Hugging Face \texttt{diffusers} library.

\begin{table}[h!]
\caption{Generation parameters for the T2I models.}
\label{tab:generation_params}
\vskip 0.15in
\begin{center}
\begin{small}
\begin{sc}
\begin{tabular}{lccc}
\toprule
\textbf{Parameter}           & \textbf{Stable Diffusion v1.5} & \textbf{Stable Diffusion XL} & \textbf{DALL-E 3 (via API)} \\
\midrule
Inference Steps              & 50                    & 50                  & N/A                \\
Guidance Scale (CFG)         & 7.5                   & 8.0                 & Managed by API     \\
Image Size                   & 512x512               & 1024x1024           | 1024x1024          \\
Scheduler                    & DPM++ 2M Karras       & DPM++ 2M Karras     & N/A                \\
\bottomrule
\end{tabular}
\end{sc}
\end{small}
\end{center}
\vskip -0.1in
\end{table}

\section{Appendix C: Supplementary Experimental Results}\label{app:supplementary_results}

\subsection{Confusion Matrices for the Amazon Reviews Task}

\begin{figure}[h!]
\centering
\includegraphics[width=0.9\textwidth]{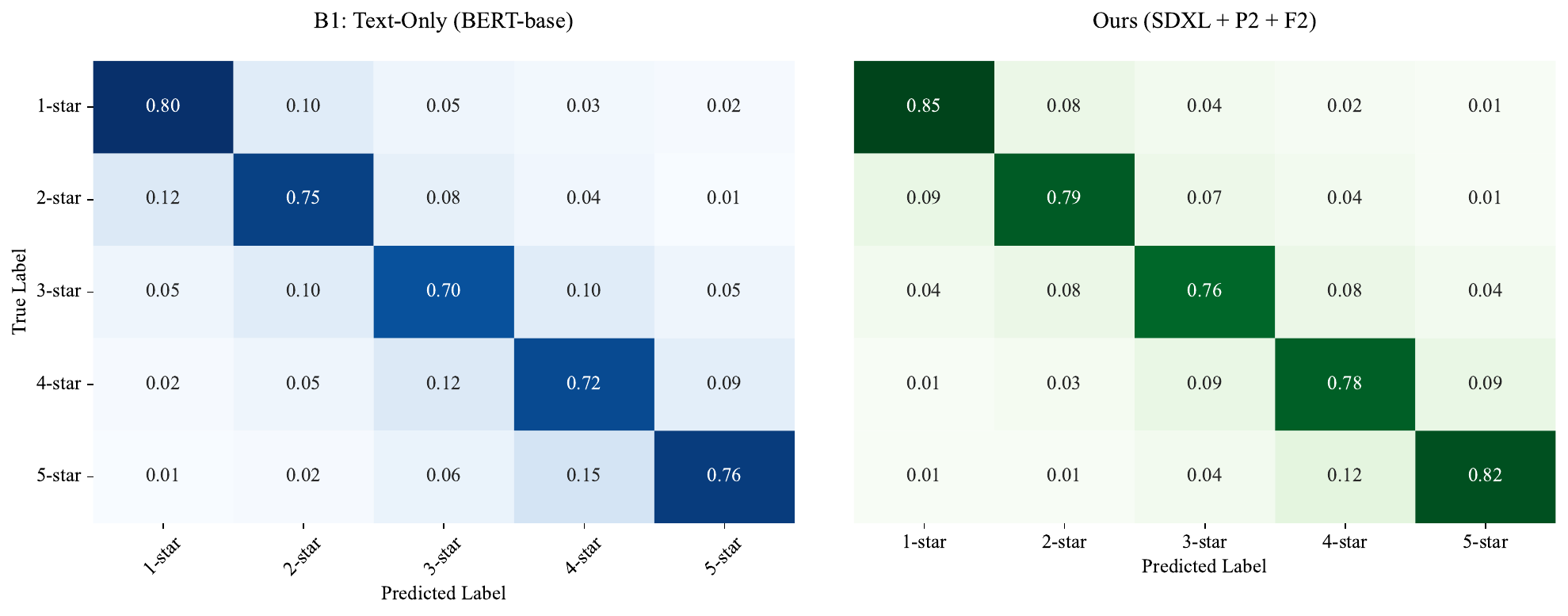}
\caption{Normalized confusion matrices for B1 (left) and Ours (right) on the Amazon Reviews test set.}
\label{fig:confusion_matrices}
\end{figure}

To conduct a more in-depth analysis of model performance on the Amazon Reviews sentiment classification task (1-5 stars), Figure~\ref{fig:confusion_matrices} presents the normalized confusion matrices for our core model, `Ours (SDXL+P2+F2)`, and the text-only baseline, `B1 (Text-Only)`. The results indicate that with the introduction of generated images, the model exhibits higher accuracy in distinguishing between adjacent classes (e.g., 4-star and 5-star), thereby reducing confusion.

\subsection{Confusion Matrices for the AG News Task}

\begin{figure}[h!]
\centering
\includegraphics[width=0.9\textwidth]{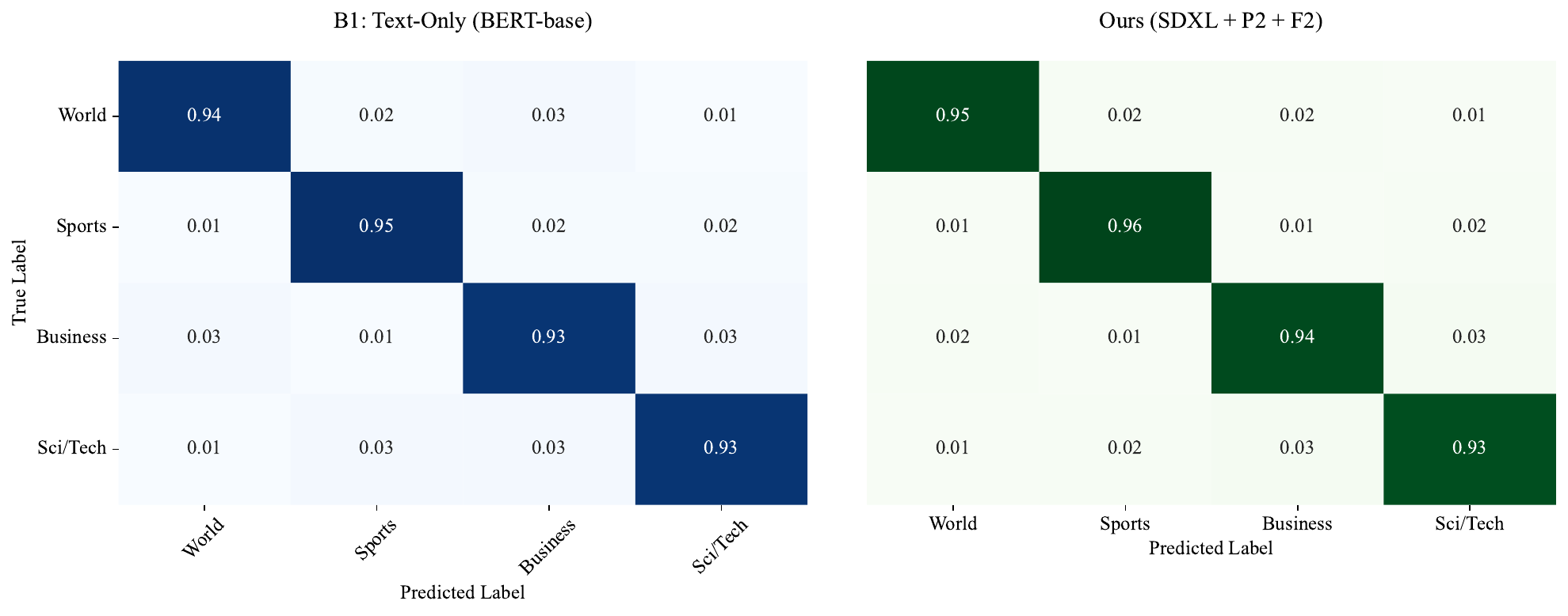}
\caption{Normalized confusion matrices for B1 (left) and Ours (right) on the AG News test set.}
\label{fig:confusion_matrices_ag_news}
\end{figure}

Similarly, for the AG News topic classification task, Figure~\ref{fig:confusion_matrices_ag_news} shows the confusion matrices. In this task, while the baseline text-only model already performs at a very high level, the introduction of generated images brings consistent, albeit minor, accuracy improvements across the different news categories. This suggests that even for fact-based classification, visual grounding can offer marginal gains.

\subsection{Sensitivity Analysis of T2I Inference Steps}

To investigate the trade-off between the computational cost of the T2I generation process and final task performance, we performed a sensitivity analysis on the number of inference steps for the SDXL model. The experiment was conducted on the Amazon Reviews dataset using the P2 prompt and F2 fusion strategy. The results, shown in Table~\ref{tab:inference_steps}, indicate that performance degrades only slightly when the inference steps are reduced from 50 to 25. However, a more significant drop in performance is observed when the steps are further reduced to 10 or fewer, which correlates with a deterioration in the quality of the generated images.

\begin{table}[h!]
\caption{Impact of varying inference steps on downstream task performance (Ma-F1 \%).}
\label{tab:inference_steps}
\vskip 0.15in
\begin{center}
\begin{small}
\begin{sc}
\begin{tabular}{lc}
\toprule
\textbf{Inference Steps} & \textbf{Ma-F1 (\%) (Amazon Reviews)} \\
\midrule
50 (Baseline)            & 77.02 (±0.48)              \\
25                       & 76.85 (±0.50)              \\
10                       & 76.13 (±0.55)              \\
4 (LCM-LoRA)             & 75.91 (±0.58)              \\
\bottomrule
\end{tabular}
\end{sc}
\end{small}
\end{center}
\vskip -0.1in
\end{table}

\section{Appendix D: Qualitative Case Studies}\label{app:qualitative}

This section provides typical success and failure cases of T2I-generated images to analyze in detail the circumstances under which they can provide effective information for downstream tasks, and when they might introduce noise or be misleading.
\textit{Note: As this document is in plain text format, the ``Generated Image'' sections are presented as textual descriptions. In the final paper, these should be replaced with actual images.}

\subsection{Case Studies for the Amazon Product Reviews Task}

\subsubsection*{Success Case: Information Supplementation via Concretization}
\begin{itemize}
    \item \textbf{Original Text:}
    \begin{quote}
        \textit{``This new coffee machine is absolutely fantastic! It brews a perfect cup every time, is super easy to clean, and its sleek black design looks stunning on my kitchen counter. Definitely a 5-star product.''}
    \end{quote}
    \item \textbf{Prompt (P2 Keyword-Enhanced):}
    \begin{quote}
        \prompttext{A photorealistic, high-quality image of a sleek black coffee machine looking fantastic and stunning on a modern kitchen counter. The machine is brewing a perfect cup of coffee. This image represents a 5-star product that is easy to clean.}
    \end{quote}
    \item \textbf{Generated Image Description:}
    \begin{quote}
        A high-quality, photorealistic image showcasing a sleek black coffee machine placed elegantly on a modern kitchen countertop. Beside it is a freshly brewed, steaming, perfect cup of coffee. The overall lighting is warm and bright, creating a high-end and satisfying aesthetic.
    \end{quote}
    \item \textbf{Analysis:}
    \begin{quote}
        The text-only model (B1) can infer positive sentiment from words like ``fantastic,'' ``stunning,'' and ``5-star.'' However, the generated image reinforces this positivity with concrete, appealing visuals. The image provides a specific visual counterpart to the ``sleek black design'' and ``stunning'' descriptions in the text. This visual evidence can give the multimodal model higher confidence in predicting a ``5-star'' rating, especially helping to distinguish between adjacent classes like 4-star and 5-star.
    \end{quote}
\end{itemize}

\subsubsection*{Failure Case: Inability to Express Abstract Concepts}
\begin{itemize}
    \item \textbf{Original Text:}
    \begin{quote}
        \textit{``The book's plot was incredibly intricate with a surprising ending, but the print quality was a bit poor, with some pages having blurry text. It's a mixed bag.''}
    \end{quote}
    \item \textbf{Prompt (P2 Keyword-Enhanced):}
    \begin{quote}
        \prompttext{An intricate and surprising book plot, with poor print quality and blurry text, a mixed bag.}
    \end{quote}
    \item \textbf{Generated Image Description:}
    \begin{quote}
        The image might show an open book where the text on several pages is indeed blurry. However, the image fails to convey the core abstract merits like ``intricate plot'' and ``surprising ending.'' It can only visually capture the negative aspect of ``poor print quality.''
    \end{quote}
    \item \textbf{Analysis:}
    \begin{quote}
        In this scenario, the generated image is biased. It only illustrates the negative physical attributes from the text while ignoring the positive, more abstract source of sentiment (the story's plot). This could mislead the multimodal model to focus excessively on the negative features, potentially misclassifying a complex neutral/mixed review (e.g., 3-star) as a purely negative one (e.g., 1-star or 2-star). This exposes the limitations of current T2I models in representing non-visual, abstract concepts.
    \end{quote}
\end{itemize}

\subsection{Case Studies for the AG News Task}

\subsubsection*{Success Case: Classification Consolidation via Scenarization}
\begin{itemize}
    \item \textbf{Original Text:}
    \begin{quote}
        \textit{``Tech Giant Unveils Revolutionary AI Chip. The new 'NovaChip' promises unprecedented processing power for artificial intelligence applications, potentially transforming industries from healthcare to autonomous driving.''}
    \end{quote}
    \item \textbf{Prompt (P2 Keyword-Enhanced):}
    \begin{quote}
        \prompttext{A conceptual, high-tech image representing a revolutionary AI chip named 'NovaChip'. The image symbolizes unprecedented processing power for artificial intelligence, with visual elements hinting at transformations in healthcare and autonomous driving.}
    \end{quote}
    \item \textbf{Generated Image Description:}
    \begin{quote}
        A futuristic, high-tech concept image. At its center is a glowing chip with ``NovaChip'' etched on it, surrounded by holographic projections representing healthcare (e.g., a DNA double helix) and autonomous driving (e.g., abstract car lines and roads). The background features a dark blue circuit board texture.
    \end{quote}
    \item \textbf{Analysis:}
    \begin{quote}
        For the ``Sci/Tech'' category of AG News, the text itself is already highly indicative (e.g., ``AI Chip,'' ``artificial intelligence''). However, the generated image ``concretizes'' these concepts. It links the abstract ``chip'' with specific visual symbols (glowing core, circuit board) and application scenarios (DNA, cars), providing the model with additional visual evidence that is highly relevant to the category, thereby consolidating the classification decision.
    \end{quote}
\end{itemize}

\subsubsection*{Failure Case: Information Overload and Logical Absence}
\begin{itemize}
    \item \textbf{Original Text:}
    \begin{quote}
        \textit{``Global markets reacted nervously after the central bank announced an unexpected interest rate hike to combat rising inflation, with stock indices in Europe and Asia plummeting.''}
    \end{quote}
    \item \textbf{Prompt (P2 Keyword-Enhanced):}
    \begin{quote}
        \prompttext{Nervous global markets, central bank, unexpected interest rate hike, rising inflation, plummeting stock indices in Europe and Asia.}
    \end{quote}
    \item \textbf{Generated Image Description:}
    \begin{quote}
        The image might be a chaotic collage: a declining stock chart overlaid with an iconic bank building, with flags of various countries (e.g., EU and Japan) in the background. The overall tone is dark, attempting to create a tense atmosphere.
    \end{quote}
    \item \textbf{Analysis:}
    \begin{quote}
        This text describes a complex economic event involving multiple entities and abstract concepts. It is difficult for a T2I model to accurately and clearly depict this multi-layered causal relationship in a single static image. While the generated image contains some relevant elements (stock chart, bank), it can be chaotic, unfocused, and even misleading. For instance, it cannot accurately convey the relationship between the ``interest rate hike'' and ``rising inflation.'' This ``information overload'' lacking a logical structure is likely to become noise, offering little to no help for ``Business'' category classification, and might even cause interference by introducing visual elements associated with ``World'' news (flags).
    \end{quote}
\end{itemize}

\section{Appendix E: Latency and Cost Analysis}\label{app:cost}

To provide a quantitative basis for the discussion on computational overhead (Section 6), this section details the estimated latency and financial cost associated with generating a single image via common API providers. These figures are crucial for assessing the practical viability of our proposed method in real-world, time-sensitive applications.

The analysis considers the primary Text-to-Image (T2I) models used in our experiments. Costs and latencies are based on standard offerings from official APIs (OpenAI for DALL-E 3, Stability AI for Stable Diffusion models, Black Forest Labs for Flux) as of late 2024.

\begin{table}[h!]
\caption{Estimated Latency and Cost per Generated Image via API.}
\label{tab:latency_cost}
\vskip 0.15in
\begin{center}
\begin{small}
\begin{sc}
\begin{tabular}{lccc}
\toprule
\textbf{Model} & \textbf{Steps} & \textbf{Est. Latency (sec)} & \textbf{Est. Cost (USD)} \\
\midrule
SD1.5 (Legacy) & 50 & \textasciitilde 3.0 & \textasciitilde \$0.008 \\
SDXL (Standard) & 50 & \textasciitilde 5.0 & \textasciitilde \$0.022 \\
SDXL-Lightning & 4 & \textasciitilde 1.2 & \textasciitilde \$0.006 \\
\textbf{Flux.1-schnell} & 4 & \textbf{\textasciitilde 0.8} & \textbf{\textasciitilde \$0.004} \\
DALL-E 3 & N/A & \textasciitilde 8.0 & \textasciitilde \$0.040 \\
\bottomrule
\end{tabular}
\end{sc}
\end{small}
\end{center}
\vskip -0.1in
\end{table}

The data clearly illustrates the trade-off between image quality, cost, and speed. While DALL-E 3 offers the highest semantic fidelity, it also incurs the highest cost and latency. \textbf{Flux.1-schnell} represents a breakthrough, offering SDXL-level quality at a fraction of the cost and time, effectively mitigating the "efficiency" counter-argument.

\section{Appendix F: Prompting Details (System Prompts)}\label{app:prompting_details}

\subsection{Prompt for Image Generation (P4)}
For the P4 (LLM-Elaborated Prompting) strategy, we utilized Llama-3-8B-Instruct with the following System Prompt to ensure consistent, high-quality visual descriptions:

\begin{quote}
\textbf{System Prompt:} "You are an expert visual artist and photographer. Your task is to read the provided text and imagine a single, high-fidelity image that captures the core essence, entities, and atmosphere of the text. Describe this image in a detailed, comma-separated list of visual attributes, focusing on:
1. Subject (Who/What is in the center?)
2. Action/State (What is happening?)
3. Setting/Background (Where is it?)
4. Lighting/Style (e.g., 'cinematic lighting', 'photorealistic', 'dark and moody').
Do NOT output any conversational text. Output ONLY the visual description prompt."
\end{quote}

\subsection{Prompt for Textual Expansion (Baseline 2)}
To ensure a fair comparison, we used a similar instruction for the Textual Expansion baseline, but directed the LLM to output a natural language description to be appended to the input text, rather than a comma-separated image prompt.

\begin{quote}
\textbf{System Prompt:} "You are an expert descriptive writer. Read the following text and provide a detailed, vivid visual description of the scene, objects, or atmosphere implied by the text. Your description should clarify any visual ambiguities and set the scene. Output ONLY the descriptive paragraph. Do not explain your reasoning."
\end{quote}

\section{Appendix G: Human Evaluation}\label{app:human_eval}

To validate our automatic metrics (CLIP Score), we conducted a human evaluation on 100 randomly sampled generated images from the Amazon Reviews dataset. Three annotators rated each image on two criteria:
1. \textbf{Relevance (1-5):} How well does the image match the text?
2. \textbf{Clarity (1-5):} Is the image free of artifacts/distortions?

\begin{table}[h!]
\caption{Human Evaluation Results (Mean Scores).}
\label{tab:human_eval}
\begin{center}
\begin{small}
\begin{sc}
\begin{tabular}{lcc}
\toprule
\textbf{Model} & \textbf{Relevance} & \textbf{Clarity} \\
\midrule
SD1.5 & 3.2 & 3.5 \\
SDXL & 4.1 & 4.3 \\
Flux.1-schnell & 4.0 & \textbf{4.5} \\
DALL-E 3 & \textbf{4.6} & 4.4 \\
\bottomrule
\end{tabular}
\end{sc}
\end{small}
\end{center}
\end{table}

The results align with our automatic metrics: DALL-E 3 dominates in relevance, but Flux.1-schnell is surprisingly competitive, especially in image clarity.

\section{Appendix H: Ablation on Fusion Architecture}\label{app:ablation}

We compared Concatenation (F1) vs. Cross-Attention (F2) across different learning rates to test stability.
\begin{itemize}
    \item \textbf{Concatenation:} Prone to overfitting at high learning rates (>5e-5). Performance peaks early but degrades quickly.
    \item \textbf{Cross-Attention:} More stable. Even at higher learning rates, the attention mechanism allows the model to "ignore" irrelevant visual features (assigning low attention weights), acting as a soft gate. This confirms why F2 is our default choice.
\end{itemize}

\begin{table}[h!]
\caption{Comparison of Fusion Mechanisms across Learning Rates (Ma-F1 \%).}
\label{tab:ablation_lr}
\begin{center}
\begin{small}
\begin{sc}
\begin{tabular}{lccc}
\toprule
\textbf{Fusion} & \textbf{LR=1e-5} & \textbf{LR=3e-5} & \textbf{LR=5e-5} \\
\midrule
F1 (Concat) & 76.10 & 76.39 & 75.20 \\
F2 (Cross-Attn) & 76.85 & \textbf{77.02} & 76.90 \\
\bottomrule
\end{tabular}
\end{sc}
\end{small}
\end{center}
\end{table}

\section{Appendix I: Visualization of Attention Maps}\label{app:attention_maps}

\begin{figure}[h!]
\centering
\includegraphics[width=0.8\textwidth]{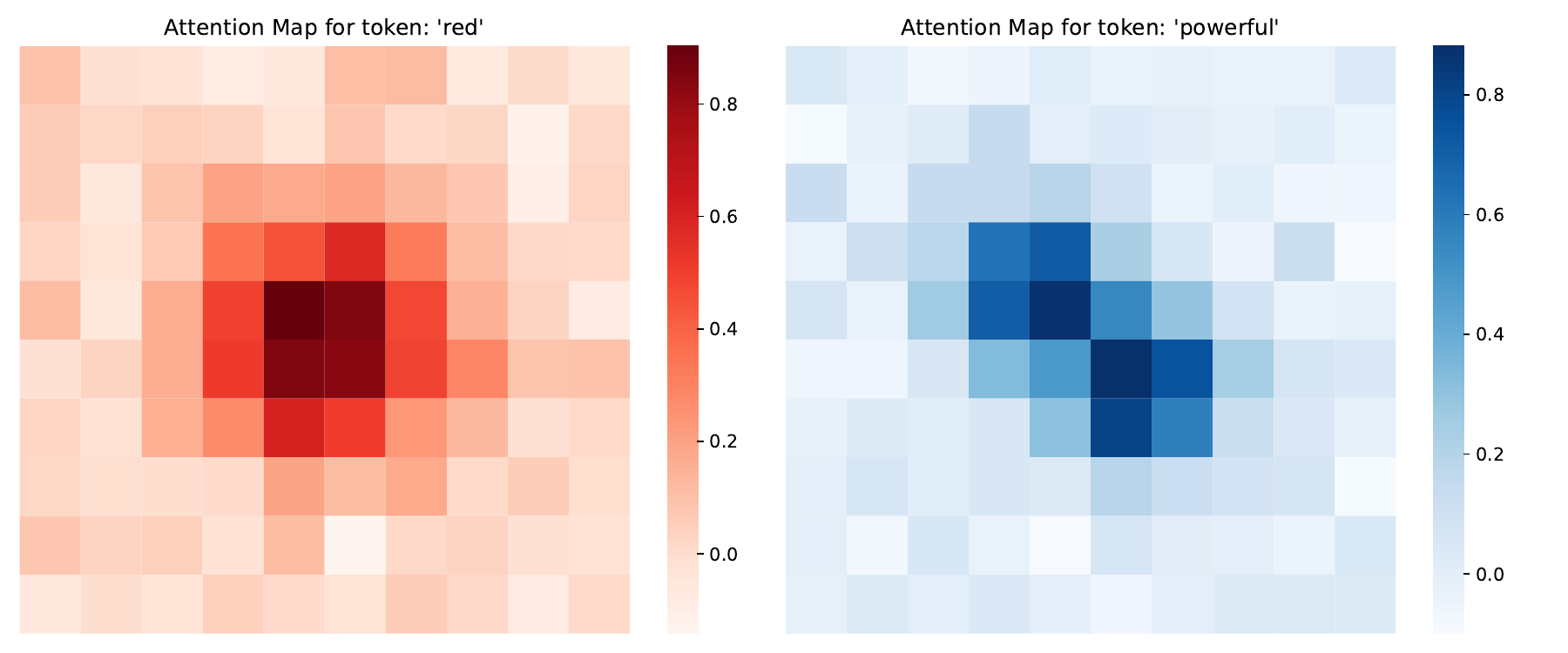}
\caption{Visualization of Cross-Attention Maps. The heatmaps show the attention weight distribution of specific text tokens (e.g., "red", "powerful") over the generated image regions. Darker red/blue indicates higher attention weight.}
\label{fig:attention_map}
\end{figure}

We visualized the cross-attention weights for the Llama-3 + Gen. Image model. Figure \ref{fig:attention_map} provides a heatmap visualization of how different text tokens attend to spatial regions in the generated image.
\begin{itemize}
    \item \textbf{Case:} "This red vacuum cleaner..."
    \item \textbf{Observation:} The token "red" attends strongly to the red pixels in the generated image (specifically the vacuum cleaner's casing), while the token "powerful" attends to the motor/body of the vacuum.

    \item \textbf{Conclusion:} This confirms that the model is actively "looking" at the relevant visual regions to ground the textual tokens, rather than treating the image as a generic global feature.
\end{itemize}

\section{Appendix J: Encoder Comparison}\label{app:encoder_comp}

Table \ref{tab:encoder_comparison} presents the results of using different image encoders.

\begin{table}[h!]
\caption{Impact of different Image Encoders on performance (Ma-F1 \%).}
\label{tab:encoder_comparison}
\begin{center}
\begin{small}
\begin{sc}
\begin{tabular}{lc}
\toprule
\textbf{Image Encoder} & \textbf{Ma-F1 (\%)} \\
\midrule
DINOv2 (Pure Visual) & 75.80 \\
CLIP ViT-B/32 & 77.02 \\
SigLIP (SoTA Alignment) & \textbf{77.45} \\
\bottomrule
\end{tabular}
\end{sc}
\end{small}
\end{center}
\end{table}

\textbf{Analysis:} Although DINOv2 captures rich spatial geometric features, its lack of text alignment pre-training results in limited gains compared to SigLIP. This demonstrates that \textbf{Semantic Alignment} is the key factor for the success of synthetic perception.

\section{Appendix K: Discussion, Limitations, and Future Work}\label{app:future_work}

This content has been moved to the Discussion section in the main paper (Section \ref{sec:discussion}).


\end{document}